\documentclass[english,onecolumn,12pt, letterpaper]{IEEEtran}
\usepackage[T1]{fontenc}
\usepackage[latin9]{inputenc}
\usepackage{geometry}
\usepackage{enumitem}
\usepackage{amsmath}
\usepackage{amsthm}
\usepackage{amssymb}
\usepackage{tikz}
\usepackage{graphicx}
\usepackage{pgfplots}
\usepackage{setspace}
\usepackage{xfrac}
\usepackage{subcaption}
\usepackage{xcolor}
\usepackage{orcidlink}

\usepackage{setspace}
\normalsize

\makeatletter
\theoremstyle{plain}
\newtheorem{thm}{\protect\theoremname}
\theoremstyle{plain}
\newtheorem{prop}[thm]{\protect\propositionname}
\theoremstyle{plain}
\newtheorem{lem}[thm]{\protect\lemmaname}
\theoremstyle{plain}

\usepackage{hyperref}
\usepackage{enumitem}
\usepackage{breqn}
\usepackage{bbm} 
\usepackage{cite}

\DeclareMathOperator*{\argmax}{arg\,max} \DeclareMathOperator*{\argmin}{arg\,min}

\allowdisplaybreaks

\global\long\def\s[#1]{\textnormal{\scriptsize #1}}
\global\long\def\st[#1]{\textnormal{\tiny #1}}

\global\long\def\la{\left(}
\global\long\def\ra{\right)}
\global\long\def\lb{\left[}
\global\long\def\rb{\right]}
\global\long\def\lc{\left\{}
\global\long\def\rc{\right\}}

\newcommand{\dfn}{\stackrel{\triangle}{=}}

\global\long\def\P{\mathbb{P}}
\global\long\def\E{\mathbb{E}}

\global\long\def\I{\mathbbm{1}}

\global\long\def\r[#1]{#1}

\global\long\def\trre[#1,#2]{\overset{{\scriptstyle (#2)}}{#1}} 

\makeatother

\usepackage{babel}
\providecommand{\lemmaname}{Lemma}
\providecommand{\propositionname}{Proposition}
\providecommand{\theoremname}{Theorem}
\providecommand{\remarkname}{Remark}

\newcommand{\calA}{{\cal A}}
\newcommand{\calB}{{\cal B}}
\newcommand{\calC}{{\cal C}}

\newcommand{\calF}{{\cal F}}

\newcommand{\calN}{{\cal N}}

\newcommand{\calP}{{\cal P}}

\newcommand{\calS}{{\cal S}}

\newcommand{\dint}{\mathrm{d}}

\newcommand {\bp} {\boldsymbol{p}}

\newcommand {\bx} {\boldsymbol{x}}
\newcommand {\by} {\boldsymbol{y}}
\newcommand {\bz} {\boldsymbol{z}}

\newcommand {\bP} {\boldsymbol{P}}
\newcommand {\bU} {\boldsymbol{U}}
\newcommand {\bV} {\boldsymbol{V}}

\newcommand {\bY} {\boldsymbol{Y}}
\newcommand {\bZ} {\boldsymbol{Z}}

\newcommand {\nn} {\nonumber}

\begin{document}

\title{Achievable Rates and Error Probability Bounds of Frequency-based Channels of Unlimited Input Resolution
\thanks{$^{1}$ R. Tamir is an independent researcher, email: \texttt{einsran@gmail.com}. $^{2}$ N. Weinberger is with the Department of Electrical and Computer Engineering, Technion, Haifa 3200003, Israel, email: \texttt{nirwein@technion.ac.il}. The research of N. Weinberger was partially supported by the Israel Science Foundation (ISF), grant no. 1782/22. This paper was accepted in part to the 2025 IEEE International Symposium on Information Theory (ISIT), Ann Arbor, Michigan, 22-27 June, 2025. 

The present work has two main contributions -- a random coding bound and an expurgated bound. The expurgated bound is not included at all in the conference version. In addition, the proof of the random coding bound is rather long, and the conference version only includes its outline, with many missing details. Finally, the current version contains additional numerical results, through discussion and comparisons.} }

\author{Ran Tamir \orcidlink{0000-0002-8291-6092}$^{1}$ and Nir Weinberger \orcidlink{0000-0001-6028-8892}$^{2}$}

\thispagestyle{empty}

\maketitle

\begin{abstract}
We consider a molecular channel, in which messages are encoded to the frequency of objects in a pool, and whose output during reading time is a noisy version of the input frequencies, as obtained by sampling with replacement from the pool. Motivated by recent DNA storage techniques, we focus on the regime in which the input resolution is unlimited. 
We propose two error probability bounds for this channel; 
the first bound is based on random coding analysis of the error probability of the maximum likelihood decoder and the second bound is derived by code expurgation techniques. 
We deduce an achievable bound on the capacity of this channel, and compare it to both the achievable bounds under limited input resolution, as well as to a converse bound.

\end{abstract}

\section{Introduction}

In the ever-increasing demand for digital archival storage, DNA storage
systems emerge as a promising technology thanks to their extremely high
information density and potential longevity \cite{church2012next,goldman2013towards,grass2015robust,yazdi2015rewritable,kiah2016codes,erlich2017dna,sala2017exact,organick2018random,lenz2019anchor,sima2021coding,tang2021error}.
In DNA storage, or, more generally, in molecular storage \cite{gohari2016information},
information is encoded via the presence of certain objects in some
restricted physical location, which we refer to as the \emph{pool}.
For example, the pool could be a tube holding DNA molecules. At reading
time, a random object is sampled from the pool and then identified. In the DNA storage case, a DNA molecule sampled from the pool is sequenced
to obtain the sequence of neucloutides it is comprised from. This
sampling and reading operation is done multiple times, to obtain a
noisy version of the objects sampled. This list does not match exactly the objects in the pool,
since the sampling is performed \emph{with replacement}, and so some
objects may be sampled more than once, while others not at all. In principle,
the identification of the object may also be noisy \cite{sabary2021solqc},
but in this paper we focus on noiseless identification.
The capacity of such storage systems is an active research area, and
recent works include, e.g. \cite{shomorony2021dna,lenz2019upper,lenz2020achieving,weinberger2022Error,weinberger2022dna,shomorony2022information,vippathalla2023secure}. 

More concretely, in the DNA storage channel model, also referred to as
the \emph{shuffling channel} \cite{shomorony2022information}, $K$
molecules are stored, each has length $L=\beta\log_{|{\cal A}|}K$
symbols from the alphabet ${\cal A}=\{\text{A},\text{C},\text{G},\text{T}\}$.
If $\beta>1$, then the index of the molecule can be
stored as a header of length $\log K=L/\beta$, and the rest $L(1-1/\beta)$
symbols can be used to encode the data. In the noiseless case we
consider, this scheme is capacity achieving \cite{shomorony2021dna}.
However, if $\beta<1$ then the indexing method fails, and in fact,
the capacity is provably zero. Hence, the log-cardinality of the optimal
codebook (which is the total number of stored bits) scales at most
\emph{sub-linearly} with the total number of nucleotides $KL$. Nonetheless,
due to the extreme information density of the DNA medium, the total
number of stored bits may be extremely large even if the capacity
is zero. In \cite{gerzon2024capacity}, a conjecture made in \cite[Sec. 7.3]{shomorony2022information}
was (partially) resolved, and the log-cardinality of the optimal codebook
was shown to increase for $\beta\in(\frac{1}{2},1)$ as 
\begin{equation}
\frac{1-\beta\log|{\cal A}|}{2\beta}\cdot K^{\beta\log|{\cal A}|}\log K+o\left(1/\log K\right).\label{eq: optimal log-cardinality DNA}
\end{equation}

The main idea of \cite{gerzon2024capacity} was to consider the regime
of $\beta<1$ as a \emph{frequency-based channel}. Indeed, in this
regime, the number of molecules in a codeword $K$ is larger than
the number of possible molecules of length $L$ from the alphabet ${\cal A}$,
to wit, $K\geq|{\cal A}|^{L}$. Accordingly, each codeword must contain
multiple copies of the same molecule (at least for one molecule). Hence,
in the short-molecule regime, the message is actually encoded by the
number of times each of the $|{\cal A}|^{L}$ possible molecules of
length $L$ appears in the pool.

To characterize the maximal log-cardinality of such a
model, a more general model was proposed, in which the blocklength $n$ of a codeword
models the number of different types of objects. Each codeword
has a total count of $ng_{n}$ objects of different types, 
which are sampled $nr_{n}$ times in total. For example, in the DNA
storage channel $n=|{\cal A}|^{L}$, as this is the number of different
molecules of length $L$ from an alphabet ${\cal A}$, and $ng_{n}=K$,
as this is the total number of molecules. Upper and lower bounds on
the capacity of this channel were derived, which after translation
to the parameters defining the DNA storage channel, lead to (\ref{eq: optimal log-cardinality DNA}).
A key aspect of the model and analysis of \cite{gerzon2024capacity}
is that up to normalization, the input is a vector of $n$ integers
that sum to $ng_{n}$. Thus, the input has \emph{finite resolution},
and this affected both the converse part, as well as the direct part,
in which the loss due to this finite resolution was bounded. The main
tactic of \cite{gerzon2024capacity} is to modify the channel model,
which follows a multinomial distribution, to a \emph{Poisson channel}
with input constraint $\mathbb{E}[X]\leq g_{n}$ and output $Z\mid X=x\sim\text{Poisson}(\frac{r_{n}}{g_{n}}x)$
(see also the related \cite{bello2024lattice}). 

In this paper, we consider the same channel model, with $n$ different
objects and random sampling with replacement, but for \emph{unlimited}
total number of objects (i.e., $g_{n}$ is effectively infinite).
The input is thus a point in an $(n-1)$-dimensional simplex, of \textit{unlimited resolution}. The
reduction to a Poisson channel in the converse bound of \cite{gerzon2024capacity}
is still applicable, and so the upper bound on the capacity of
$\frac{1}{2}\log(r_{n})+o_{n}(1)$ is intact. By contrast, the reduction
technique in the direct part is based on certain asymptotic relations
between $r_{n}$ and $g_{n}$, which do not apply for infinite $g_{n}$.
We therefore take a direct approach in this paper and develop an
achievable bound for this channel as a function of $r_{n}$, the normalized
number of samples. 
As a first step towards proving an achievable bound, we propose two error probability bounds for this channel; 
the first bound is based on random coding analysis of the error probability of the maximum likelihood (ML) decoder and the second bound stems from code expurgation, in the spirit of the classic technique \cite[Sec.\ 5.7]{gallager1968information}. While the bound related to code expurgation strictly improves upon the random coding probability bound at low coding rates, the random coding bound holds for a wider range of coding rates, especially high rate. Thus, we use the random coding bound to deduce a single-letter expression for an achievable rate.

Our model is related to the  \textit{multinomial channel}, explored classically in the context of universal compression \cite{rissanen1984universal,davisson1973universal}. In this channel model, the alphabet $\cal A$ has finite cardinality, and a single input to the channel is a point in the $(|{\cal A}|-1)$-dimensional
simplex, representing a probability distribution over $\cal A$. The channel output is a multinomial distribution, whose event distribution is determined by the input and the number of trials is $nr_n$.
When considering multiple, independent, uses of this channel, the capacity of the resulting memoryless channel is then achieved, as usual, by long codewords with such inputs, and was derived in \cite{xie1997minimax,clarke1994jeffreys}. In the context of DNA storage, this channel model is relevant to \textit{composite DNA
storage} \cite{choi2019high,anavy2019data}, in which each stored input is an arbitrary mixture of the letters in ${\cal A}=\{\text{A},\text{C},\text{G},\text{T}\}$.  During reading, each composite letter is sampled multiple
times, and this allows us to estimate the input mixture. The resulting
channel is the multinomial channel, whose capacity was explored, e.g.,
in \cite{chang1990numerical, komninakis2001capacity,farsad2020capacities,kobovich2023m}. 

In contrast,  our setting assumes a \textit{single} use of a channel,\footnote{We use $n$ as a scaling parameter to determine asymptotics. In other words, if one consider the $n$ input frequencies as $n$ inputs, the resulting channel is not memoryless (it is not even obvious if it is information stable).} and that the input-alphabet cardinality scales linearly with the input length $n$. Therefore, we do not rely on a single-letter mutual information term to describe the capactity, but rather derive error probability bounds from first principles, and then provide an achievable rate as the maximal rate in which our error probability bound decays to zero. 

Our model is indeed relevant to practical DNA storage: one of the reasons that the total number of molecules is bounded by
$K$ is to control synthesis costs. However, the actual sequencing
cost is for \emph{different} molecules, since once a molecule is synthesized,
the costs of duplicating it are relatively low. As recently proposed in \cite{preuss2024efficient}, the mixture in the composite DNA method could
be of short molecules (shortmers), rather than just single nucleotides, and so the dimensionality
of the simplex can be larger than $3$. The regime we consider in this paper is the one in
which the constituent alphabet size is $n$, and thus proportional to
the number of samples $nr_{n}$. We also mention the closely related
\emph{permutation channel} \cite{kovavcevic2017coding,kovavcevic2018codes,makur2020coding,tang2023capacity,lu2024permutation},
in which each input object is sampled exactly once, though in a random order. 

The paper is organized as follows. In Section \ref{Section2} we formulate our model and state our main objective. In Section \ref{Section3} we introduce and discuss the main result of this work and compare it to the main results of \cite{gerzon2024capacity}. 
The proofs of the main results are provided in Sections \ref{Section4} and \ref{Section5}. The paper is concluded in Section \ref{Section6}.


\section{Problem Formulation} \label{Section2}

\subsection{Codebook generation}
Consider a set of $n$ distinguishable types of substances. The channel input is represented by the mixture vector $\bp^n :=(p_1,\ldots,p_n)$, where $p_i$ is the concentration of the $i$th type in the pool of substances. 
Thus, each codeword $\bp^n$ is a probability mass function (PMF) in the $n$-dimensional simplex $\calP_n = \{(p_1,\ldots,p_n)\mid p_i \geq 0, \sum_{i=1}^n p_i =1 \}$. In general, one can sample a random PMF $\bP^n=(P_1,\ldots,P_n)$ from $\calP_n$ using the $n$-dimensional Dirichlet distribution with parameters $(\alpha_1,\ldots,\alpha_n)$.  
First, draw $n$ independent random samples $(S_1,\ldots,S_n)$ from Gamma distributions each with density
\begin{align}
f_{S_i}(s_i)
=\text{Gamma}(\alpha_i,1)
=\frac{s_i^{\alpha_i-1}e^{-s_i}}{\Gamma(\alpha_i)},
\end{align}
and then set
\begin{align}
P_i = \frac{S_i}{\sum_{j=1}^{n}S_j}.
\end{align}
In this work, we study the symmetric case $\alpha_1=\ldots=\alpha_n=\alpha$ for some $\alpha>0$ and construct a random codebook $\calC_n$ by independently drawing $M$ codewords $\{\bP^n(1),\ldots,\bP^n(M)\}$ using the described method.  

\subsection{Encoding and Decoding}

For a given message $m$ uniformly drawn from $\{1,2,\ldots,M\}$, one pours the $n$ distinct substances according to the concentrations given by the PMF $\bp^n(m)$. When reading the message, one samples the pool independently $nr_n$ times (with replacement), and the type of each sample is read by a noiseless mechanism. 


We denote the random samples by $\by=\{y_1,\ldots,y_{nr_n}\}$, where $y_i \in [n]$ for each $i \in [nr_n]$, which follow a multinomial distribution with $nr_n$ trials and parameters $\bp^n(m)$. The optimal ML decoder calculates the enumerators
\begin{align}
N_{\by}(\ell) := \sum_{i=1}^{nr_n} \I\{y_i = \ell\},
\end{align}
then decodes the message according to 
\begin{align} \label{Decoder_ML}
\hat{m}_{\mbox{\tiny ML}}(\by) =\argmax_{m \in \{1,\ldots,M\}} \prod_{\ell=1}^{n} p_{\ell}(m)^{N_{\by}(\ell)}.
\end{align}
One can easily show that the ML decoder is equivalent to a decoder that estimates the message as the one whose codeword minimizes the Kullback--Leibler (KL) divergence with $\hat{P}_{\by} = (\hat{P}_{\by}(1),\ldots,\hat{P}_{\by}(n))$, where $\hat{P}_{\by}(\ell) := N_{\by}(\ell)/(nr_n)$.
Indeed, 
\begin{align}
\hat{m}(\by)
&=\argmin_{m \in \{1,\ldots,M\}} D(\hat{P}_{\by}\|\bp^{n}(m)) \\
&=\argmin_{m \in \{1,\ldots,M\}} \left[\sum_{\ell=1}^{n} \hat{P}_{\by}(\ell) \log \hat{P}_{\by}(\ell) - \sum_{\ell=1}^{n} \hat{P}_{\by}(\ell) \log p_{\ell}(m) \right]  \\
&=\argmax_{m \in \{1,\ldots,M\}}  \sum_{\ell=1}^{n} nr_n\hat{P}_{\by}(\ell) \log p_{\ell}(m) \\
&=\argmax_{m \in \{1,\ldots,M\}}  \prod_{\ell=1}^{n} p_{\ell}(m)^{nr_n\hat{P}_{\by}(\ell)},
\end{align}
which is easily identified as the ML decoder.

The probability of error of the ML decoder is given by
\begin{align}
    \varepsilon_n = \P\lb \hat{m}(\bY) \neq m \rb,
\end{align}
which is taken with respect to the randomness of the message selection, the random codebook generation, and the sampling process.

The size of the largest code for $n$ substance types, normalized number of samples $r_n$, 
under a given error probability $\varepsilon_n \in (0,1)$ is denoted by $M^*(\varepsilon_n, r_n)$. 
Our goal is to accurately determine the growth rate of $M^*(\varepsilon_n, r_n)$, or the rate of the codebook, given by $\frac{1}{n} \log M^*(\cdot)$, under the condition that $\varepsilon_n \to 0$ as $n \to \infty$.
We assume that $r_n$ is a monotonically non-decreasing function of $n$, and aim to accurately characterize the dependency of $M^*(\varepsilon_n, r_n)$ on this sequence. 
In addition, assuming that $M=e^{nR}$ and $R \leq \frac{1}{n} \log M^*(\cdot)$, we would like to characterize the decay rate of $\varepsilon_n$ as a function of $R$.

\section{Main Results} \label{Section3}

\subsection{A Random Coding Exponent}

We begin with our first result, which is a random coding bound. Before presenting this bound, we make some definitions. The gamma function is defined as
\begin{align}
\Gamma(z) := \int_{0}^{\infty} t^{z-1} e^{-t} \dint t, 
\end{align}
and Riemann's zeta function by
\begin{align}
\zeta(s) := \sum_{n=1}^{\infty} \frac{1}{n^{s}}.
\end{align}
In addition, let us denote
\begin{align}
\Psi(t) 
:= (1+t) H_{2}\la \frac{1}{1+t} \ra = (1+t) \log(1+t) - t\log t,
\end{align}
where $H_{2}(p) := -p\log(p)-(1-p)\log(1-p)$ is the binary entropy function. We also define the functions 
\begin{align}
\label{Def_Theta}
\Lambda(r_n, \alpha, \xi)
&:= \alpha \Psi\left(\frac{\xi r_n}{\alpha}\right) - \frac{(2\alpha-1)}{2}\log \left(\alpha + \xi r_n \right) - \frac{1}{2}\log (2\pi) + \log\Gamma(\alpha), \\
\Delta(r_n)
&:=  \inf_{q>2} \left\{ \left(1-\frac{1}{q}\right) \Psi(r_n) + \frac{1}{q}\log\left[1 + (2\pi)^{-q/2} \zeta\left(\frac{q}{2}\right) \right] \right\} ,
\end{align}
and the \textit{exponent function} 
\begin{align} \label{exponent_function}
E_{\mbox{\scriptsize r}}(R,r_n) 
&:= \sup_{\alpha > 1/2} \sup_{\xi > 0} \sup_{\mu > 0} \min\left\{ \lb \Lambda(r_n, \alpha, \xi) - \xi(\Delta(r_n) + \mu) - R \rb_{+}, \mu \right\}.
\end{align}

\begin{thm} \label{Theorem_main_Probability}
Assume that $\{r_n\}_{n \geq 1}$ is a sequence that grows sub-exponentially fast. Then,
\begin{align} \label{Lower_bound}
    \varepsilon_n \leq 2\sqrt{2\pi e nr_n}  \exp \left\{ -n \cdot E_{\mbox{\scriptsize r}}(R,r_n) \right\}.
\end{align}
\end{thm}
The form of the exponent function in \eqref{rate_function} can be explained as follows: Before maximizing over $\mu, \xi,$ and $\alpha$, the exponent function $E_{\mbox{\scriptsize r}}(R,r_n)$ is given by the minimum between two expressions. This follows since the overall error event is effectively given by the union of two disjoint events. The first error event is standard, and is related to the errors that stem from the $e^{nR}-1$ incorrect codewords that are packed in the same space together with the true codeword, and each one of them may be closer, in principle, to the statistics of the $nr_n$ samples.   
The second error event is related to the sampling process. The $nr_n$ samples that are drawn from the true PMF $\bp$, which is the codeword representing the message, typically have  a composition that is closer (in the KL divergence sense) to the true $\bp$ more than any other codeword. However, this rarely may not be the case, and the resulting composition of the samples may be closer to some incorrect PMF codeword, causing a decoding error.
In Appendix  \ref{appendix_lemma_proof}  we prove the following result, which asserts that this is indeed a large deviation event. 

\begin{lem} \label{LD_lemma}
Let $n,r_n \in \mathbb{N}$, $\bp \in \calP_{n}$, and $\bZ \sim \mathrm{Multinomial}(nr_n,\bp)$. Then, for any $\mu > 0$,
\begin{align} \label{LD_equation}
\P\left[ D\left(\frac{\bZ}{nr_n}\middle\|\bp\right) \geq \frac{\Delta(r_n) + \mu}{r_n} \right]
\leq \sqrt{2\pi e nr_n} e^{-n\mu}.
\end{align} 
\end{lem}

We conclude from Lemma \ref{LD_lemma} that with a probability close to one, $D\left(\frac{\bZ}{nr_n}\middle\|\bp\right) \leq \frac{\Delta(r_n)}{r_n}$, and since $\Delta(r_n)$ grows logarithmically fast in $r_n$, this means that for large $r_n$, the sample statistics are close to $\bp$ with high probability. 

We pause to mention that the probabilities of moderate- and large- deviations related to multinomial distributions, very similar to \eqref{LD_equation}, were extensively studied in \cite{kallenberg1985moderate}. However, it is important to note that its results are not easily adopted for our needs. 
The results of \cite{kallenberg1985moderate} assume a deterministic $\bp$, whereas we need further averaging over $\bp$, which is a realization of $\bP \sim \text{Dir}(\alpha)$. The probability bound in \cite[Theorem 2.1]{kallenberg1985moderate} is given as a relatively complicated function of $\bp$, and averaging it with respect to the Dirichlet distribution  is challenging. Furthermore, the large-deviations result in \cite[Corollary 2.4]{kallenberg1985moderate} is valid under some conditions related to $\bp$, which are relatively difficult to verify in our setting, again, due to the fact that $\bp$ itself is a realization of a random PMF. 
Finally, the bound in \cite[Eq.\ (2.6)]{kallenberg1985moderate} is the only one which does not depend on $\bp$; however, it is easy to check that this bound is loose compared to \eqref{LD_equation}.    

\subsection{An Expurgated Exponent}

We next turn to our second bound, the expurgated exponent. 
It turns out that the error probability bound given in Theorem \ref{Theorem_main_Probability} may be improved at low coding rates by code expurgation. In this case, unlike the random coding bound in Theorem \ref{Theorem_main_Probability}, which holds on average, over the ensemble of random codebooks, the expurgated bound only assures the existence of at least one code with exceptionally good qualities.

For a given code $\calC_{n}$, the probability of error given that message $m$ was transmitted is given by
\begin{align}
P_{e|m}(\calC_{n}) = \sum_{\by \in [n]^{nr_n}} P(\by|\bp_{m}) \I\left\{\hat{m}(\by)\neq m\right\},
\end{align}
where $P(\by|\bp_{m})$ stands for the probability of sampling the vector $\by$, given that the codeword PMF is $\bp_{m}$. 
Before presenting the result, we make the following definitions:
\begin{align}
F(\kappa) &:= \frac{4}{\sqrt{2\pi}} \int_{0}^{\infty}\dint x  \exp\left\{-\frac{1}{2}(1-\kappa^2) x^2 \right\} \Phi(\kappa x) \\
G(\lambda \sigma) &:= \sup_{\kappa \in (0,1)} [\kappa \lambda \sigma - \log F(\kappa)] \\
J(\sigma) &:= \frac{1}{2}[\sigma - \log(\sigma) - 1] \\
L(\lambda) &:= \sup_{\sigma \in (0,1)} \min\{G(\lambda \sigma), J(\sigma)\} \\
S(r_n,\rho) &:= \sup_{\lambda \in (0,1)} \min\left[\frac{r_n}{\rho}\log\left(\frac{1}{\lambda}\right), L(\lambda)\right].
\end{align}

\begin{thm} \label{Theorem_Expurgated}
There exists a sequence of codes, $\{\calC_{n}\}_{n\geq 1}$, such that 
\begin{align}
\liminf_{n \to \infty} -\frac{1}{n} \log \max_{m} P_{e|m}(\calC_{n}) \geq E_{\mbox{\scriptsize ex}}(R,r_n),
\end{align}
where
\begin{align}
E_{\mbox{\scriptsize ex}}(R,r_n) = \sup_{\rho >1} \{\rho \cdot (S(r_n,\rho)-R)\}.
\end{align}
\end{thm}

To prove the existence of a good code as in Theorem \ref{Theorem_Expurgated}, we start with a random codebook, composed by PMF codewords drawn independently from $\text{Dir}(\frac{1}{2})$. Then, we show probabilistically that there exists at least one codebook, such that after expurgating one half of its codewords, the remaining codewords become sufficiently far apart to induce relatively low conditional error probabilities.   

We briefly explain why, in principle, the bound in Theorem \ref{Theorem_Expurgated} may be improved.
At the beginning of the proof of Theorem \ref{Theorem_Expurgated} in Section \ref{Section5}, we show that 
\begin{align}
P_{e|m}(\calC_{n})
\leq \sum_{m' \neq m} \left(\sum_{k=1}^{n} \sqrt{p_{m}(k) p_{m'}(k)} \right)^{nr_n} 
\dfn \sum_{m' \neq m} 
\text{BC}(\bp_{m}, \bp_{m'})^{nr_n},
\end{align}
where $\bp_{m} = (p_{m}(1),\ldots,p_{m}(n))$ is the true codeword and $\bp_{m'} = (p_{m'}(1),\ldots,p_{m'}(n))$ is a competing codeword. 
From this point on, deriving the tightest bound requires one to know the moments\footnote{Or, alternatively, the behavior of $\P\left[ \text{BC}(\bP_{m}, \bP_{m'}) \geq \lambda \right]$, where $\lambda \in (0,1)$.} of $\text{BC}(\bP_{m}, \bP_{m'})$ for any $\alpha>0$. Although this happens to be a complicated probabilistic task for a general $\alpha>0$, fortunately enough, these moments can be assessed for $\alpha=\frac{1}{2}$, thanks to the representation of the Dirichlet distribution as in \eqref{Representation} below, which involves normal random variables that are relatively easy to handle. 
However, a probabilistic characterization of $\text{BC}(\bP_{m}, \bP_{m'})$ for any $\alpha>0$ may yield a tighter bound. This point is suggested for future research.        

\subsection{Discussion}
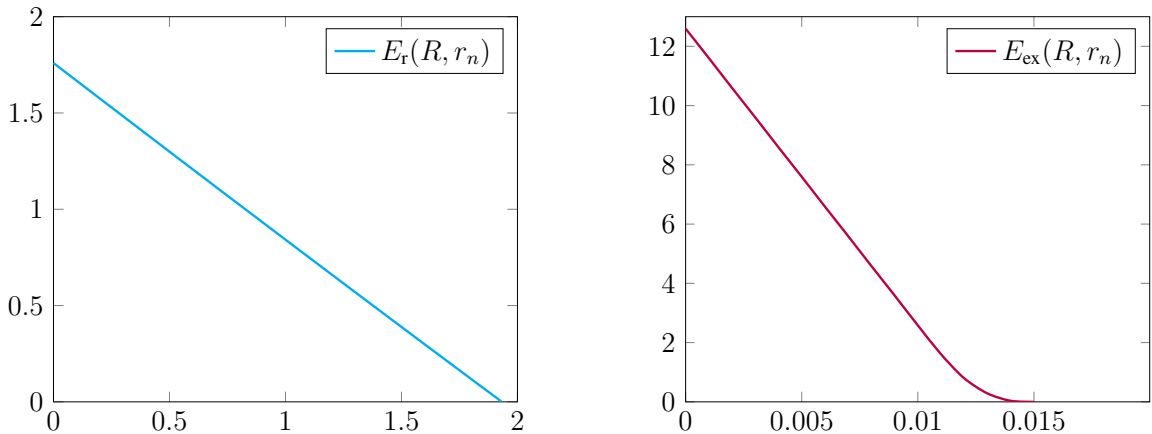
\begin{figure}[h!]	
	\begin{subfigure}[b]{0.5\columnwidth}
		\centering 
\begin{tikzpicture}[scale=0.9]
	\begin{axis}[
	disabledatascaling,
	scaled x ticks=false,
	xticklabel style={/pgf/number format/fixed,
		/pgf/number format/precision=3},
	scaled y ticks=false,
	yticklabel style={/pgf/number format/fixed,
		/pgf/number format/precision=3},
	xmin=0, xmax=2,
	ymin=0, ymax=2,
	legend pos=north east
	]

\addplot[color=cyan,line width=1pt,smooth]
table {
0.0 1.7595 
0.05 1.7134 
0.1 1.6672 
0.15 1.6211 
0.2 1.5751 
0.25 1.529 
0.3 1.483 
0.35 1.4371 
0.4 1.3911 
0.45 1.3452 
0.5 1.2993 
0.55 1.2535 
0.6 1.2077 
0.65 1.1619 
0.7 1.116 
0.75 1.0703 
0.8 1.0247 
0.85 0.9791 
0.9 0.9335 
0.95 0.8879 
1.0 0.8423 
1.05 0.7968 
1.1 0.7514 
1.15 0.706 
1.2 0.6605 
1.25 0.6152 
1.3 0.57 
1.35 0.5246 
1.4 0.4794 
1.45 0.4341 
1.5 0.389 
1.55 0.344 
1.6 0.2989 
1.65 0.254 
1.7 0.2089 
1.75 0.164 
1.8 0.1191 
1.85 0.0742 
1.9 0.0296 
1.93 0.0026
};
\legend{}
\addlegendentry{$E_{\mbox{\scriptsize r}}(R,r_n)$}

	\end{axis}
	\end{tikzpicture}
		\label{fig:Random_Coding}
	\end{subfigure}
	\begin{subfigure}[b]{0.5\columnwidth}
		\centering 
\begin{tikzpicture}[scale=0.9]
	\begin{axis}[
	disabledatascaling,
	scaled x ticks=false,
	xticklabel style={/pgf/number format/fixed,
		/pgf/number format/precision=3},
	scaled y ticks=false,
	yticklabel style={/pgf/number format/fixed,
		/pgf/number format/precision=3},
	xmin=0, xmax=0.02,
	ymin=0, ymax=13,
	legend pos=north east
	]

\addplot[color=purple,line width=1pt,smooth]
table {
0.0 12.5969 
0.001 11.5969 
0.002 10.5969 
0.003 9.5969 
0.004 8.5969 
0.005 7.5969 
0.006 6.5969 
0.007 5.5969 
0.008 4.5969 
0.009 3.5969 
0.01 2.5969 
0.011 1.6141 
0.012 0.8055 
0.013 0.2876 
0.014 0.0357 
0.015 -0.0005
};
\legend{}
\addlegendentry{$E_{\mbox{\scriptsize ex}}(R,r_n)$}

	\end{axis}
	\end{tikzpicture}		
		\label{fig:Expurgated}
	\end{subfigure}
	\caption{The exponent functions $E_{\mbox{\scriptsize r}}(R,r_n)$ and $E_{\mbox{\scriptsize ex}}(R,r_n)$ as functions of $R$ for $r_n=400$.} 
	\label{fig:Exponent}
\end{figure}

We begin our discussion with a numerical example. Figure \ref{fig:Exponent} shows $E_{\mbox{\scriptsize r}}(R,r_n)$ and $E_{\mbox{\scriptsize ex}}(R,r_n)$ as functions of $R$ for $r_n=400$. As can be seen, the exponent function that stems from code expurgation improves upon the random coding error exponent at relatively low coding rates. Since the shapes of these exponent functions are rather unusual, a few words are in order. First, $E_{\mbox{\scriptsize r}}(R,r_n)$ has no strictly convex part at relatively high rates. Based on the results in \cite{kallenberg1985moderate}, we speculate that by further tightening Lemma \ref{LD_lemma}, the improved exponent function may present a curvy part, and leave this point for future work.     
Second, $E_{\mbox{\scriptsize ex}}(R,r_n)$ is expected to be strictly convex at rates approaching zero, which is not the case as presented in Figure \ref{fig:Exponent}. In this case, the affine shape at low rates seems to be a numerical artifact, since we numerically solve the optimization problem with $\rho \in [1,\rho_{0}]$, where theoretically $\rho>1$. Note that the optimal $\rho^{*}(R)$ is related to the local gradient of the curve at rate $R$, then restricting this parameter from above is the reason for the affine shape found in Figure \ref{fig:Exponent}.    

From the error probability bound provided in Theorem \ref{Theorem_main_Probability} one can easily deduce a lower bound on achievable rates. A rate $R$ is achievable as long as the exponent function $E_{\mbox{\scriptsize r}}(R,r_n)$ is strictly positive.
Let us define the rate function 
\begin{align} \label{rate_function}
R_{\mbox{\tiny LB}}(r_n) 
&:= \sup_{\alpha > 1/2} \sup_{\xi > 0} \left\{ \Lambda(r_n, \alpha, \xi) - \xi \Delta(r_n)  \right\}.
\end{align}
\begin{thm} \label{Theorem_main}
Assume that $\{r_n\}_{n \geq 1}$ is a sequence that grows sub-exponentially fast. Then,
\begin{align}
    \frac{1}{n}\log M^*(\varepsilon_n, r_n) \geq R_{\mbox{\tiny LB}}(r_n).
\end{align}
\end{thm}

We would like to stress that the capacity is strictly positive, and that exponentially many messages can be encoded in the $n$-dimensional simplex $\calP_n$ should not be taken for granted for the following reason. Unlike ordinary coding problems, for example the sphere-packing problem for the additive Gaussian channel, where the average power of the transmitted codeword grows linearly with block length, in our setting, the power of the ``strongest'' codeword is $1$ (on the corner points of the simplex), regardless of $n$. 
Still, we are able to transmit with positive rates, as the decoder is allowed to have an effective channel output sequence (the $nr_n$ samples) which may be much longer than $n$.

We next discuss our choice of input distribution. In this paper, we assume that the codewords $\{\bP^n(1),\ldots,\bP^n(M)\}$ are drawn from a symmetric Dirichlet distribution over the $(n-1)$-dimensional simplex $\calP_n$, with parameters $\alpha_1=\ldots=\alpha_n=\alpha$ such that $\alpha > 0$. For technical reasons related to our bounding techniques (specifically, the passages to \eqref{ref11} and \eqref{ref13}), our bound of achievability holds only for $\alpha\geq \frac{1}{2}$. Furthermore, optimizing numerically over $\alpha$ reveals that the optimizer is $\alpha^{*}=\frac{1}{2}$, regardless of $r_n$. We suspect that the optimal $\alpha$ may be even closer to zero, for the simple reason that drawing PMFs with a relatively low $\alpha$ ``pushes'' most of them to the boundary of the simplex and, as a consequence, causes them to be more distant from each other (in the KL divergence sense).      
Strengthening Theorem \ref{Theorem_main} to provide an achievability bound that allows for any $\alpha>0$ is left for future research.

We next justify our choice of the Dirichlet distribution as the random coding distribution over other distributions defined on the $(n-1)$-dimensional simplex. In principle, a general methodology to generate random points on the simplex is as follows. Take any positive random variable $X \sim P$ and draw $n$ independent copies $X_1,\ldots,X_n \sim P$. Now, the random vector defined by
\begin{align} \label{Representation}
(U_1,\ldots,U_n) := \left(\frac{X_{1}}{\sum_{i=1}^{n}X_{i}},\ldots,\frac{X_{n}}{\sum_{i=1}^{n}X_{i}} \right)
\end{align}
is obviously supported on the $(n-1)$-dimensional simplex.
It is well known that the Dirichlet distribution is equivalent to the representation in \eqref{Representation} for some special cases. E.g., if $X_1,\ldots,X_n \sim \text{Exp}(1)$, then $(U_1,\ldots,U_n)$ follows the Dirichlet distribution with $\alpha=1$, which, in turn, is equivalent to the uniform distribution over $\calP_n$. In addition, if $Z_1,\ldots,Z_n \sim \calN(0,1)$ and $X_{i} := Z_{i}^{2}$ for any $i \in [n]$, then $(U_1,\ldots,U_n)$ follows the Dirichlet distribution with $\alpha=\frac{1}{2}$.

As we show in the sequel, the Dirichlet distribution has a special characteristic which makes it a perfect match to the optimal ML decoder, which is based on the KL divergence.
At the beginning of the proof of Theorem \ref{Theorem_main_Probability}, when handling the pairwise error probability $\P[ D(\hat{P}_{\by}\|\bP) \leq D(\hat{P}_{\by}\|\bp)]$, where $\bp$ is the true codeword, $\bP$ is a competing codeword, and $\hat{P}_{\by}$ is the empirical distribution of the vector of samples, we upper-bound this probability using Chernoff's inequality, and then, in the next step, we need to handle an expectation of the form
\begin{align} \label{Expectation}
\E\left[\prod_{i=1}^{n} P_i^{\theta\hat{P}_{\by}(i)} \right],    
\end{align}
where $\theta>0$ is a parameter. 
As opposed to products of independent random variables, where their expectations can be calculated by pulling the multiplication operation outside, when dealing with products of dependent random variables, things are usually not so easy, to say the least. In some cases, one may rely on special characterizations of the random variables at hand, e.g., negative association, which allows\footnote{See, for example, \cite[p.\ 288, Property $\text{P}_2$]{joag1983negative}.} (possibly under some monotonicity conditions) to bound the expectation of the product by the product of individual expectations. Unfortunately, this may lead to loose bounds in many cases.
Although the various components of the vector $\bP=(P_1,\ldots,P_n)$ are statistically dependent, it turns out that the product moment in \eqref{Expectation} can be precisely evaluated when $\bP$ follows a Dirichlet distribution. The following result concerning product moments of the Dirichlet distribution can be found, e.g., in \cite[p.\ 274]{balakrishnan2004primer}. 

\begin{prop} \label{Prop_Dirichlet}
Let $(\alpha_1,\ldots,\alpha_n)$ and $(\beta_1,\ldots,\beta_n)$ be positive vectors and let $(X_1,\ldots,X_n) \sim \mathrm{Dir}(\alpha_1,\ldots,\alpha_n)$. Then, it holds that
\begin{align}
\label{Dirichlet_Moments}
\E\lb \prod_{i=1}^{n} X_i^{\beta_i} \rb
= \frac{\Gamma\la \sum_{i=1}^{n} \alpha_i \ra}{\Gamma\la \sum_{i=1}^{n} (\alpha_i+\beta_i) \ra} \cdot
\prod_{i=1}^{n} \frac{\Gamma\la \alpha_i+\beta_i \ra}{\Gamma\la \alpha_i \ra}.
\end{align}
\end{prop}


As can be easily seen, the expectation in \eqref{Expectation} can be evaluated exactly using the result of Proposition \ref{Prop_Dirichlet}, and as a consequence, we now see that, at least from a technical point of view, the Dirichlet distribution land itself naturally as a matching random coding distribution for the ML decoder in the studied problem.

We next compare our main result with those recently derived in \cite{gerzon2024capacity} for the case of finite input resolution of the possible inputs in the $n$-dimensional simplex. More specifically, the channel model in \cite{gerzon2024capacity} assumes that $n$ types of objects are available, and the channel input is a count vector $\bx := (x_1,\ldots,x_n) \in \mathbb{N}^n$, where $x_i$ is the number of objects of the $i$-th type in the pool of objects. It is assumed that $\sum_{i=1}^{n}x_i$ is a constant for all possible messages, and grows as $ng_n$, where $g_n<\infty$. The reading is done in the same manner as our model. In  \cite{gerzon2024capacity} lower and upper bounds on the best achievable information rate (defined similarly as above) were provided. Specifically, their lower bound is given by the expression (neglecting  asymptotically vanishing terms) 
\begin{align}
R_{\mbox{\tiny LB}}^{\mbox{\tiny FIR}}(g_n,r_n) = \frac{1}{2}\log(r_n) - \Psi\left(\frac{r_n}{g_n}\right).
\end{align}
For a fixed $g_n$, it was shown in \cite{gerzon2024capacity} that $R_{\mbox{\tiny LB}}^{\mbox{\tiny FIR}}(g_n,r_n)$ has a unique global maximum around $r_n \approx 0.398g_n$, which means that for finite input resolution, a finite sampling rate is optimal. 

\begin{figure}[h!]
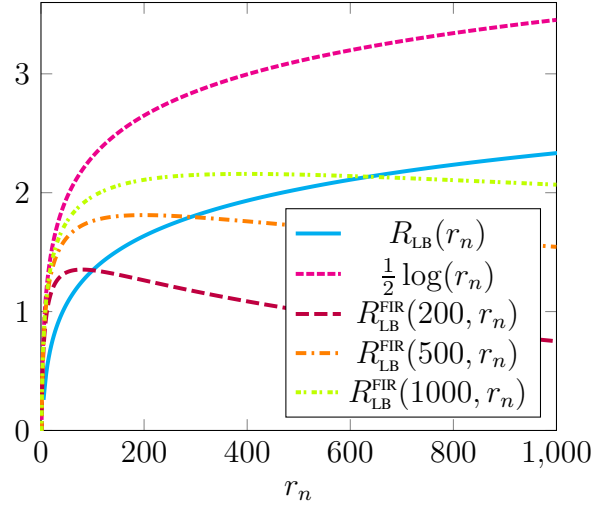
	
	\centering 
 
	\caption{A comparison between $R_{\mbox{\tiny LB}}(r_n)$, $R_{\mbox{\tiny LB}}^{\mbox{\tiny FIR}}(g_n,r_n)$ for three $g_n$ values, and the converse expression $\frac{1}{2}\log(r_n)$.} 
	\label{fig:LowerBound}
\end{figure}

Figure \ref{fig:LowerBound} compares our rate $R_{\mbox{\tiny LB}}(r_n)$ to the rate  $R_{\mbox{\tiny LB}}^{\mbox{\tiny FIR}}(g_n,r_n)$ of \cite{gerzon2024capacity} as a  function of $r_n$ for three values of $g_n$.
We stress, however, that the curves shown for $R_{\mbox{\tiny LB}}^{\mbox{\tiny FIR}}(g_n,r_n)$ may not be valid for all values of $r_n$, and so the comparison is delicate. The figure should thus be interpreted with caution. 
As can be seen in Figure \ref{fig:LowerBound}, the curves of $R_{\mbox{\tiny LB}}^{\mbox{\tiny FIR}}(g_n,r_n)$ have global maxima, while the curve of $R_{\mbox{\tiny LB}}(r_n)$ increases without bound. We also plot the converse bound $\frac{1}{2}\log(r_n)$ (without the $o_n(1)$ term), which is valid in our setting too, and can be proved by repeating the steps of the proof of the converse part of \cite[Theorem 2]{gerzon2024capacity} almost verbatim. Now, for some $g_n$, the rate $R_{\mbox{\tiny LB}}^{\mbox{\tiny FIR}}(g_n,r_n)$ may appear larger than $R_{\mbox{\tiny LB}}(r_n)$ for low values of $r_n$ (and approach the upper bound when $g_n$ is increased). However, it must be emphasized that the curves $R_{\mbox{\tiny LB}}^{\mbox{\tiny FIR}}(g_n,r_n)$ are only valid under the assumptions $g_n=\Theta(r_n)$ and $n=\omega(g_n)$ of \cite[Theorem 2]{gerzon2024capacity}. Thus, for any given $r_n$ (and $n$), the bound of \cite[Theorem 2]{gerzon2024capacity} does not specify exactly how large $g_n$  can be. For example, while for $r_n=200$ the finite-resolution rate for $g_n=1000$ is larger than our bound on the infinite-resolution rate, it is not assured by  
\cite[Theorem 2]{gerzon2024capacity} that this rate is achievable for such a large $g_n$. In contrast, our bound does not have this restriction at all, and applies for any $n$. For large values of $r_n$ our random coding bound becomes larger than any finite-resolution bound of \cite{gerzon2024capacity} for a given $g_n$, as expected.

\section{Proof of Theorem \ref{Theorem_main_Probability}} \label{Section4}
We assume without loss of generality that the true message is $m=1$. 
Let us condition on the codeword $\bP^n(1)=\bp=(p_1,\ldots,p_n)$ and on the channel output sequence $\bY=\by$ and denote $\bP^n=\bP^n(m)=(P_1,\ldots,P_n)$.
The probability of error is given by 
\begin{align}
\varepsilon_n(\bp,\by)
&= \P\lb \bigcup_{m=2}^{M} \{D(\hat{P}_{\by}\|\bP^n(m)) \leq D(\hat{P}_{\by}\|\bp)\} \rb \\
\label{ref0}
&\leq \min \lc 1, \sum_{m=2}^{M} \P\lb D(\hat{P}_{\by}\|\bP^n(m)) \leq D(\hat{P}_{\by}\|\bp) \rb \rc, 
\end{align}
using the clipped union bound. 

Let $\theta \geq 0$ be an arbitrary parameter.
The probability in \eqref{ref0} is given by
\begin{align}
\P\left[ D(\hat{P}_{\by}\|\bP^n) \leq D(\hat{P}_{\by}\|\bp) \right]
&= \P\lb \sum_{i=1}^{n} \hat{P}_{\by}(i) \log \frac{\hat{P}_{\by}(i)}{P_i} \leq \sum_{i=1}^{n} \hat{P}_{\by}(i) \log \frac{\hat{P}_{\by}(i)}{p_i} \rb \\
&= \P\lb \sum_{i=1}^{n} \hat{P}_{\by}(i) \log P_i \geq \sum_{i=1}^{n} \hat{P}_{\by}(i) \log p_i \rb \\
&= \P\lb \sum_{i=1}^{n} \log P_i^{\theta\hat{P}_{\by}(i)} \geq \theta \sum_{i=1}^{n} \hat{P}_{\by}(i) \log p_i \rb \\
&= \P\lb \prod_{i=1}^{n} P_i^{\theta\hat{P}_{\by}(i)} \geq \exp \lc \theta \sum_{i=1}^{n} \hat{P}_{\by}(i) \log p_i \rc \rb \\
\label{ref3}
&\leq \frac{\E\lb \prod_{i=1}^{n} P_i^{\theta\hat{P}_{\by}(i)} \rb}{\exp \lc \theta \sum_{i=1}^{n} \hat{P}_{\by}(i) \log p_i \rc},
\end{align}
where \eqref{ref3} follows from Markov's inequality. In order to evaluate the expectation in \eqref{ref3}, we use Proposition \ref{Prop_Dirichlet} from Section \ref{Section3}.
In our case $\alpha_1=\ldots=\alpha_n=\alpha$ and $\beta_i = \theta\hat{P}_{\by}(i)$, which yields that 
\begin{align}
\E\left[\prod_{i=1}^{n} P_i^{\theta\hat{P}_{\by}(i)} \right]
&= \frac{\Gamma\la \alpha n \ra}{\Gamma\la \sum_{i=1}^{n} (\alpha+\theta\hat{P}_{\by}(i)) \ra} \cdot
\prod_{i=1}^{n} \frac{\Gamma\la \alpha+\theta\hat{P}_{\by}(i) \ra}{\Gamma\la \alpha \ra} \\
&= \frac{\Gamma\la \alpha n \ra}{\Gamma\la \alpha n+\theta \ra \Gamma^n(\alpha)} \cdot
\prod_{i=1}^{n} \Gamma\la \alpha+\theta\hat{P}_{\by}(i) \ra \\
\label{ref4}
&= \frac{\Gamma\la \alpha n \ra}{\Gamma\la \alpha n+\theta \ra \Gamma^n(\alpha)} \cdot
\exp \lc \sum_{i=1}^{n} \log \Gamma\la \alpha+\theta\hat{P}_{\by}(i) \ra \rc.
\end{align}
For any $t>0$, it holds from Stirling's bound that \cite[Theorem 5]{gordon1994stochastic}   
\begin{align} \label{Gamma_Bounds}
\sqrt{2\pi}t^{t-1/2}e^{-t} \leq \Gamma(t) \leq \sqrt{2\pi}t^{t-1/2}e^{-t}e^{\frac{1}{12t}},
\end{align}
and thus,
\begin{align}
&\log\Gamma(\alpha+x) \nn \\
&\leq (\alpha+x)\log(\alpha+x) - (\alpha+x) +\frac{1}{2}\log\frac{1}{\alpha+x} +\frac{1}{2}\log(2\pi) + \frac{1}{12(\alpha+x)} \\
&= x\log(\alpha+x) - (\alpha+x) +\frac{2\alpha}{2}\log(\alpha+x) - \frac{1}{2}\log(\alpha+x) +\frac{1}{2}\log(2\pi) + \frac{1}{12(\alpha+x)} \\
&= x\log(\alpha+x) - (\alpha+x) +\frac{2\alpha-1}{2}\log(\alpha+x) + \frac{1}{2}\log(2\pi) + \frac{1}{12(\alpha+x)} \\
\label{ref11}
&\leq x\log(x) - x +\frac{2\alpha-1}{2}\log(\alpha+x) + \frac{1}{2}\log(2\pi),
\end{align}
where \eqref{ref11} holds as long as $\alpha \geq \frac{1}{\sqrt{6}}$.
To justify \eqref{ref11}, consider the function
\begin{align}
f(x) = x\log(x) - x\log(\alpha+x) + \alpha - \frac{1}{12(\alpha+x)}  
\end{align}
for any $x\geq 0$. This function is continuous and it holds that 
\begin{align}
f(0)= 0\log(0) - 0\log(\alpha) + \alpha - \frac{1}{12\alpha} = \alpha - \frac{1}{12\alpha}, 
\end{align}
which is non-negative as long as $\alpha \geq \frac{1}{2\sqrt{3}}$. The derivative of $f(x)$ is bounded as
\begin{align}
f'(x) 
&= \log\left(\frac{x}{\alpha+x}\right) + \frac{\alpha}{\alpha+x} + \frac{1}{12(\alpha+x)^2}\\
&\leq -\frac{\alpha}{\alpha+x} - \frac{1}{2}\frac{\alpha^2}{(\alpha+x)^2} + \frac{\alpha}{\alpha+x} + \frac{1}{12(\alpha+x)^2}\\
&=  \frac{\frac{1}{12}-\frac{1}{2}\alpha^2}{(\alpha+x)^2},
\end{align}
which is non-positive for any $\alpha \geq \frac{1}{\sqrt{6}}$.
In addition, it follows from L'Hospital's rule that 
\begin{align}
\lim_{x \to \infty} f(x)
= \lim_{x \to \infty} \left[ x\log\left(\frac{x}{\alpha+x}\right) + \alpha - \frac{1}{12(\alpha+x)} \right] = -\alpha + \alpha -0 = 0.
\end{align}
Hence, if follows that $f(x)\geq 0$ for any $x\geq 0$ and $\alpha \geq \frac{1}{\sqrt{6}}$, which proves the passage to \eqref{ref11}.

Let us choose $\theta = \xi nr_n$ for $\xi \geq 0$ which is arbitrary at this point.
Moving forward, the sum in \eqref{ref4} is bounded as follows:
\begin{align}
\sum_{i=1}^{n} \log \Gamma\la \alpha+\xi nr_n \hat{P}_{\by}(i) \ra 
\label{ref12}
&\leq \sum_{i=1}^{n} \left[ \xi nr_n \hat{P}_{\by}(i) \log  \la \xi nr_n \hat{P}_{\by}(i) \ra - \xi nr_n \hat{P}_{\by}(i) \right. \nn \\
&~~~~\left. + \frac{2\alpha-1}{2}\log \left(\alpha+ \xi nr_n \hat{P}_{\by}(i)\right) + \frac{1}{2}\log (2\pi)   \right]  \\
\label{ref13}
&\leq  \xi nr_n \log  \la \xi nr_n \ra + \xi nr_n \sum_{i=1}^{n} \hat{P}_{\by}(i) \log \hat{P}_{\by}(i) - \xi nr_n \nn \\
&~~~~+ \frac{n(2\alpha-1)}{2}\log \left(\alpha+ \frac{1}{n} \sum_{i=1}^{n} \xi nr_n \hat{P}_{\by}(i)\right) + \frac{n}{2}\log (2\pi) \\
\label{ref5}
&=  \xi nr_n \log  \la \xi nr_n \ra + \xi nr_n \sum_{i=1}^{n} \hat{P}_{\by}(i) \log \hat{P}_{\by}(i) - \xi nr_n \nn \\
&~~~~+ \frac{n(2\alpha-1)}{2}\log \left(\alpha + \xi r_n \right) + \frac{n}{2}\log (2\pi), 
\end{align}
where \eqref{ref12} is due to \eqref{ref11} and \eqref{ref13} follows from Jensen's inequality and the concavity of the logarithmic function, which holds as long as $\alpha \geq \frac{1}{2}$, such that the factor $2\alpha-1$ is non-negative.

Substituting \eqref{ref5} back into \eqref{ref4} and then into \eqref{ref3} yields
\begin{align}
&\P\lb D(\hat{P}_{\by}\|\bP^n) \leq D(\hat{P}_{\by}\|\bp) \rb \nn \\
&\leq \frac{\Gamma\la \alpha n \ra}{\Gamma\la \alpha n+\xi nr_n \ra \Gamma^n(\alpha)} \cdot \frac{1}{\exp \lc \xi nr_n \sum_{i=1}^{n} \hat{P}_{\by}(i) \log p_i \rc} \nn \\
&~~~~\times
\exp \left\{  \xi nr_n \log  \la \xi nr_n \ra + \xi nr_n \sum_{i=1}^{n} \hat{P}_{\by}(i) \log \hat{P}_{\by}(i) - \xi nr_n \right. \nn \\
&~~~~~~\left. + \frac{n(2\alpha-1)}{2}\log \left(\alpha + \xi r_n \right) + \frac{n}{2}\log (2\pi) \right\} \\
&= \frac{\Gamma\la \alpha n \ra}{\Gamma\la \alpha n+\xi nr_n \ra} \cdot
\exp \left\{ \xi nr_n \log  \la \xi nr_n \ra + \xi nr_n D(\hat{P}_{\by}\|\bp) - \xi nr_n \right. \nn \\
&~~~~~~\left. + \frac{n(2\alpha-1)}{2}\log \left(\alpha + \xi r_n \right) + \frac{n}{2}\log (2\pi) - n \log\Gamma(\alpha) \right\} \\
\label{ref6}
&= \frac{\Gamma\la \alpha n \ra (\xi nr_n)^{\xi nr_n} \cdot \exp\lc - \xi nr_n \rc}{\Gamma\la \alpha n+\xi nr_n \ra} \nn \\
&~~\times
\exp \left\{ \xi nr_n D(\hat{P}_{\by}\|\bp) + \frac{n(2\alpha-1)}{2}\log \left(\alpha + \xi r_n \right) + \frac{n}{2}\log (2\pi) - n \log\Gamma(\alpha) \right\}.
\end{align}
It follows by the bounds in \eqref{Gamma_Bounds} that
\begin{align}
&\frac{\Gamma\la \alpha n \ra (\xi nr_n)^{\xi nr_n} \cdot \exp\lc- \xi nr_n \rc}{\Gamma\la \alpha n+\xi nr_n \ra} \nn \\
&~~~~\leq \frac{\sqrt{\frac{2\pi}{\alpha n}}e^{\frac{1}{12\alpha n}}}{\sqrt{\frac{2\pi}{\alpha n + \xi nr_n}}}
\exp \lc -n(\alpha+\xi r_n) H_{2}\la \frac{\alpha n}{\alpha n+\xi nr_n} \ra \rc \\
\label{ref16}
&~~~~= \sqrt{\frac{\alpha+\xi r_n}{\alpha}} \exp\left\{\frac{1}{12\alpha n}\right\} \exp \lc -n\alpha\left(1+\frac{\xi r_n}{\alpha}\right) H_{2}\la \frac{1}{1+\frac{\xi r_n}{\alpha}} \ra \rc \\
\label{ref15}
&~~~~\leq 2\sqrt{1+2\xi r_n} 
\exp \lc -n \alpha \Psi\left(\frac{\xi r_n}{\alpha}\right) \rc \\
\label{ref80}
&~~~~\dfn C_n \exp \lc -n \alpha \Psi\left(\frac{\xi r_n}{\alpha}\right) \rc,
\end{align}
where \eqref{ref15} is due to the facts that $\alpha \geq \frac{1}{2}$ and $n \geq 1$.

Upper-bounding \eqref{ref6} with \eqref{ref80} yields
\begin{align}
&\P\lb D(\hat{P}_{\by}\|\bP^n) \leq D(\hat{P}_{\by}\|\bp) \rb \nn \\
&\leq C_n
\exp \lc -n \alpha \Psi\left(\frac{\xi r_n}{\alpha}\right) \rc \nn \\
&~~\times
\exp \lc  \xi nr_n D(\hat{P}_{\by}\|\bp)  + \frac{n(2\alpha-1)}{2}\log \left(\alpha + \xi r_n \right) + \frac{n}{2}\log (2\pi) - n \log\Gamma(\alpha) \rc \\
&= C_n
\exp \left\{ \xi nr_n D(\hat{P}_{\by}\|\bp) - n \alpha \Psi\left(\frac{\xi r_n}{\alpha}\right) \right. \nn \\
&~~~~~~\left. + \frac{n(2\alpha-1)}{2}\log \left(\alpha + \xi r_n \right) + \frac{n}{2}\log (2\pi) - n \log\Gamma(\alpha) \right\} \\
\label{ref8} 
&= C_n
\exp \lc \xi nr_n D(\hat{P}_{\by}\|\bp) -n \Lambda(r_n, \alpha, \xi) \rc,
\end{align}
where \eqref{ref8} follows from the definition in \eqref{Def_Theta}.
Upper-bounding \eqref{ref0} with \eqref{ref8}, we arrive at
\begin{align}  
\varepsilon_n(\bp,\by)
&\leq \min \lc 1, \sum_{m=2}^{M} C_n \exp \lc -n \Lambda(r_n, \alpha, \xi) + \xi nr_n D(\hat{P}_{\by}\|\bp) \rc \rc \\
&\leq \min \lc 1, C_n \exp \lc nR -n \Lambda(r_n, \alpha, \xi) + \xi nr_n D(\hat{P}_{\by}\|\bp) \rc \rc.
\end{align}

Given the codeword $\bP^n(1)=\bp=(p_1,\ldots,p_n)$, we take the expectation with respect to the sequence $\bY$. 
Let us define the random variable $\bZ = (Z_1,\ldots,Z_n) \sim \text{Multinomial}(nr_n,\bp)$, denote its support by
\begin{align}
\calS_n := \left\{ (z_1,\ldots,z_n) \middle| \sum_{i=1}^{n} z_i = nr_n, z_i \geq 0~ \forall i \in [n] \right\},
\end{align}
and note that $\hat{P}_{\by}(i)$ that was calculated from the vector of $nr_n$ samples is also equal to $\frac{z_i}{nr_n}$.
For a given realization $\bZ=\bz$, the error probability is thus bounded as 
\begin{align}  
\varepsilon_n(\bp,\bz)
&\leq \min\left\{ 1, C_n \exp \left\{ nR -n \Lambda(r_n, \alpha, \xi) + \xi nr_n D\left(\frac{\bz}{nr_n}\middle\|\bp\right) \right\} \right\}.
\end{align}

Let us now choose 
\begin{align}
\label{Def_RHO}
\rho_n = \frac{\Delta(r_n) + \mu}{r_n}
\end{align}
and denote the set
\begin{align}
\calA_n := \left\{ \bz \in \calS_n:~ 0\leq D\left(\frac{\bz}{nr_n}\middle\|\bp\right) \leq \rho_n  \right\}.
\end{align}
To this end,
\begin{align}
\varepsilon_n(\bp)
&\leq \sum_{\bz \in \calS_n} P_{\bZ}(\bz) \min\left\{ 1, C_n \exp \left\{ nR -n \Lambda(r_n, \alpha, \xi) + \xi nr_n D\left(\frac{\bz}{nr_n}\middle\|\bp\right) \right\} \right\} \\
&= \sum_{\bz \in \calA_n} P_{\bZ}(\bz) \min\left\{ 1, C_n \exp \left\{ nR -n \Lambda(r_n, \alpha, \xi) + \xi nr_n D\left(\frac{\bz}{nr_n}\middle\|\bp\right) \right\} \right\} \nonumber \\
&~~~~+ \sum_{\bz \in \calA_n^{\mbox{\scriptsize c}}} P_{\bZ}(\bz) \min\left\{ 1, C_n \exp \left\{ nR -n \Lambda(r_n, \alpha, \xi) + \xi nr_n D\left(\frac{\bz}{nr_n}\middle\|\bp\right) \right\} \right\} \\
&\leq \sum_{\bz \in \calA_n} P_{\bZ}(\bz) C_n \exp \lc -n \cdot \lb \Lambda(r_n, \alpha, \xi) - \xi r_n \rho_n - R \rb_{+} \rc + \sum_{\bz \in \calA_n^{\mbox{\scriptsize c}}} P_{\bZ}(\bz) \\
&= \P\lb \bZ \in \calA_n \rb \cdot C_n \exp \lc -n \cdot \lb \Lambda(r_n, \alpha, \xi) - \xi r_n \rho_n - R \rb_{+} \rc + \P\lb \bZ \in \calA_n^{\mbox{\scriptsize c}} \rb \\
\label{ref1}
&\leq C_n \exp \lc -n \cdot \lb \Lambda(r_n, \alpha, \xi) - \xi r_n \rho_n - R \rb_{+} \rc + \P\lb D\left(\frac{\bZ}{nr_n}\middle\|\bp\right) \geq \rho_n \rb.
\end{align}

Continuing with $\rho_n$ as defined in \eqref{Def_RHO} provides
\begin{align}
\varepsilon_n(\bp)
&\leq 2\sqrt{1+2\xi r_n} \exp \lc -n \cdot \lb \Lambda(r_n, \alpha, \xi) - \xi \Delta(r_n) - \xi\mu - R \rb_{+} \rc \nn \\ 
&~~~~+ \P\lb D\left(\frac{\bZ}{nr_n}\middle\|\bp\right) \geq \frac{\Delta(r_n) + \mu}{r_n} \rb \\
\label{ref85}
&\leq 2\sqrt{1+2\xi r_n} \exp \lc -n \cdot \lb \Lambda(r_n, \alpha, \xi) - \xi \Delta(r_n) - \xi \mu - R \rb_{+} \rc + \sqrt{2\pi e nr_n} e^{-n\mu} \\
&\leq \sqrt{2\pi e nr_n} \left( \exp \lc -n \cdot \lb \Lambda(r_n, \alpha, \xi) - \xi \Delta(r_n) - \xi \mu - R \rb_{+} \rc + e^{-n\mu} \right) \\
&\leq 2\sqrt{2\pi e nr_n} \max\left\{ \exp \lc -n \cdot \lb \Lambda(r_n, \alpha, \xi) - \xi \Delta(r_n) - \xi \mu - R \rb_{+} \rc , e^{-n\mu} \right\} \\
\label{ref28}
&= 2\sqrt{2\pi e nr_n}  \exp \lc -n \cdot \min\left\{ \lb \Lambda(r_n, \alpha, \xi) - \xi \Delta(r_n) - \xi \mu - R \rb_{+}, \mu \right\} \rc ,
\end{align}
where \eqref{ref85} is due to Lemma \ref{LD_lemma} (in Section \ref{Section3}). Note that \eqref{ref28} is independent of the realization of $\bP^n(1)$, hence
\begin{align}
\label{Final_Expression}
\varepsilon_n 
= \E[\varepsilon_n(\bP^n(1))] 
\leq 2\sqrt{2\pi e nr_n}  \exp \left\{ -n \cdot \min\left\{ \lb \Lambda(r_n, \alpha, \xi) - \xi \cdot \Delta(r_n) - \xi \mu - R \rb_{+}, \mu \right\} \right\}, 
\end{align}
and note that the exponent function in \eqref{Final_Expression} may be maximized over the parameters $\mu,\xi$, and $\alpha$, which completes the proof of Theorem \ref{Theorem_main_Probability}.

\section{Proof of Theorem \ref{Theorem_Expurgated}}
\label{Section5}
We follow Gallager's expurgation technique \cite[Sec. 5.7]{gallager1968information}.  The main technical part of this technique is obtaining an upper bound (hopefully tight) on a fractional moment of the error probability, to wit $\E[P_{e|m}^{1/\rho}]$ for $\rho>1$. 
To this end, we begin with the standard Bhattacharyya bound \cite[Sec. 5.3]{gallager1968information} \cite[Sec. 2.3]{viterbi2009principles}: 
For any $m \in \{1,2,\ldots,M\}$, the probability of error given that message $m$ was transmitted is bounded as
\begin{align}
P_{e|m}
&= \sum_{\by \in [n]^{nr_n}} P(\by|\bp_{m}) \I\left\{\hat{m}(\by)\neq m\right\} \\
&\leq \sum_{\by \in [n]^{nr_n}} P(\by|\bp_m) \sum_{m' \neq m} \sqrt{\frac{P(\by|\bp_{m'})}{P(\by|\bp_{m})}} \\
&= \sum_{\by \in [n]^{nr_n}} \sum_{m' \neq m} \sqrt{P(\by|\bp_m)P(\by|\bp_{m'})} \\
&= \sum_{m' \neq m} \sum_{\by \in [n]^{nr_n}} \prod_{\ell=1}^{n} \sqrt{p_{m}(\ell)p_{m'}(\ell)}^{N_{\by}(\ell)}  \\
&= \sum_{m' \neq m} \sum_{\by \in [n]^{nr_n}} \prod_{\ell=1}^{n} \prod_{i=1}^{nr_n} \sqrt{p_{m}(\ell) p_{m'}(\ell)}^{\I\{y_{i}=\ell\}}  \\
&= \sum_{m' \neq m} \sum_{\by \in [n]^{nr_n}} \prod_{i=1}^{nr_n} \prod_{\ell=1}^{n} \sqrt{p_{m}(\ell) p_{m'}(\ell)}^{\I\{y_{i}=\ell\}}  \\
&= \sum_{m' \neq m} \sum_{\by \in [n]^{nr_n}} \prod_{i=1}^{nr_n} \sqrt{p_{m}(y_{i}) p_{m'}(y_{i})}  \\
&= \sum_{m' \neq m} \prod_{i=1}^{nr_n} \sum_{y_i=1}^{n} \sqrt{p_{m}(y_{i}) p_{m'}(y_{i})}  \\
&= \sum_{m' \neq m} \left(\sum_{k=1}^{n} \sqrt{p_{m}(k) p_{m'}(k)} \right)^{nr_n}  \\
&\dfn \sum_{m' \neq m} 
\text{BC}(\bp_{m}, \bp_{m'})^{nr_n},
\end{align}
where $\text{BC}(\bp,\bp') \in [0,1]$ is the Bhattacharyya coefficient between two PMFs. For $\rho \geq 1$,
\begin{align}
\E\left[P_{e|m}^{1/\rho}\right] 
&\leq \E\left[\left( \sum_{m' \neq m} 
\text{BC}(\bP_{m}, \bP_{m'})^{nr_n} \right)^{1/\rho}\right] \\
&\leq \E\left[ \sum_{m' \neq m} 
\text{BC}(\bP_{m}, \bP_{m'})^{nr_n/\rho} \right] \\
\label{ref93}
&= \sum_{m' \neq m} \E\left[ 
\text{BC}(\bP_{m}, \bP_{m'})^{nr_n/\rho} \right].
\end{align}
Let $\lambda \in (0,1)$. Then,
\begin{align}
&\E\left[ 
\text{BC}(\bP_{m}, \bP_{m'})^{nr_n/\rho} \right] \nn \\
&= \E\left[ 
\text{BC}(\bP_{m}, \bP_{m'})^{nr_n/\rho}\I\left\{ \text{BC}(\bP_{m}, \bP_{m'}) \in [0,\lambda] \right\} \right] \nn \\
&~~~~~~+ \E\left[ 
\text{BC}(\bP_{m}, \bP_{m'})^{nr_n/\rho}\I\left\{ \text{BC}(\bP_{m}, \bP_{m'}) \in [\lambda,1] \right\} \right] \\
&\leq \E\left[ 
\lambda^{nr_n/\rho}\I\left\{ \text{BC}(\bP_{m}, \bP_{m'}) \in [0,\lambda] \right\} \right] + \E\left[ 
1^{nr_n/\rho}\I\left\{ \text{BC}(\bP_{m}, \bP_{m'}) \in [\lambda,1] \right\} \right] \\
\label{ref90}
&\leq  
\lambda^{nr_n/\rho} + \P\left[ \text{BC}(\bP_{m}, \bP_{m'}) \geq \lambda \right].
\end{align}
In order to evaluate the probability in \eqref{ref90}, we confine ourselves to $\alpha=\frac{1}{2}$, and then we can use the representation in \eqref{Representation}. Let us denote the independent random variables $X_1,\ldots,X_n \sim \calN(0,1)$ and $Y_1,\ldots,Y_n \sim \calN(0,1)$, and define
\begin{align} 
\bU = (U_1,\ldots,U_n) &:= \left(\frac{X_{1}^2}{\sum_{i=1}^{n}X_{i}^2},\ldots,\frac{X_{n}^2}{\sum_{i=1}^{n}X_{i}^2} \right) \\
\bV = (V_1,\ldots,V_n) &:= \left(\frac{Y_{1}^2}{\sum_{i=1}^{n}Y_{i}^2},\ldots,\frac{Y_{n}^2}{\sum_{i=1}^{n}Y_{i}^2} \right).
\end{align}
It follows that $\bU$ and $\bV$ are independent and that $\bU, \bV \sim \text{Dir}(\frac{1}{2})$. Now, 
\begin{align}
\P\left[ \text{BC}(\bU, \bV) \geq \lambda \right]
&= \P\left[ \sum_{k=1}^{n} \sqrt{U_{k} V_{k}} \geq \lambda \right] \\
&= \P\left[ \sum_{k=1}^{n} \sqrt{\frac{X_{k}^2}{\sum_{i=1}^{n}X_{i}^2} \frac{Y_{k}^2}{\sum_{i=1}^{n}Y_{i}^2}} \geq \lambda \right] \\
&= \P\left[ \sum_{k=1}^{n} \left| \frac{X_{k}}{\sqrt{\sum_{i=1}^{n}X_{i}^2}} \frac{Y_{k}}{\sqrt{\sum_{i=1}^{n}Y_{i}^2}} \right| \geq \lambda \right] \\
&= \P\left[ \sum_{k=1}^{n} 
\left|X_{k} Y_{k} \right| \geq \lambda \sqrt{\sum_{i=1}^{n}X_{i}^2} \sqrt{\sum_{i=1}^{n}Y_{i}^2} \right].
\end{align}
For $\sigma \in (0,1)$, define the events 
\begin{align}
\calA_n(\sigma) &:= \left\{\sum_{i=1}^{n}X_{i}^2 \geq n\sigma  \right\} \\
\calB_n(\sigma) &:= \left\{ \sum_{i=1}^{n}Y_{i}^2 \geq n\sigma  \right\}
\end{align}
as well as $\calF_n(\sigma) := \calA_n(\sigma) \cap \calB_n(\sigma)$.
Then,
\begin{align}
\P\left[ \text{BC}(\bU, \bV) \geq \lambda \right]
&\leq \P\left[ \sum_{k=1}^{n} 
\left|X_{k} Y_{k} \right| \geq \lambda \sqrt{\sum_{i=1}^{n}X_{i}^2} \sqrt{\sum_{i=1}^{n}Y_{i}^2}, \calF_n(\sigma) \right] + \P\left[ \calF_n^{\mbox{\scriptsize c}}(\sigma) \right] \\
&\leq \P\left[ \sum_{k=1}^{n} 
\left|X_{k} Y_{k} \right| \geq \lambda \sqrt{n\sigma} \sqrt{n\sigma}, \calF_n(\sigma) \right] + \P\left[ \calF_n^{\mbox{\scriptsize c}}(\sigma) \right] \\
\label{ref91}
&\leq \P\left[ \sum_{k=1}^{n} 
\left|X_{k} Y_{k} \right| \geq n\lambda \sigma \right] + \P\left[ \calF_n^{\mbox{\scriptsize c}}(\sigma) \right].
\end{align}
Let $\kappa>0$. For the left-hand probability in \eqref{ref91}, 
\begin{align}
\P\left[ \sum_{k=1}^{n} 
\left|X_{k} Y_{k} \right| \geq n\lambda \sigma \right] 
&= \P\left[\exp\left\{\kappa \sum_{k=1}^{n} 
\left|X_{k} Y_{k} \right|\right\} \geq e^{n\kappa \lambda \sigma} \right] \\
&\leq e^{-n\kappa \lambda \sigma} \cdot \E\left[\prod_{k=1}^{n} \exp\left\{\kappa 
\left|X_{k} Y_{k} \right|\right\} \right] \\
\label{ref95}
&= e^{-n\kappa \lambda \sigma} \cdot \prod_{k=1}^{n} \E\left[\exp\left\{\kappa 
\left|X_{k} Y_{k} \right|\right\} \right].
\end{align}
Let $X$ and $Y$ be independent with $X,Y \sim \calN(0,1)$. Then, the expectation in \eqref{ref95} is derived as follows:
\begin{align}
\E\left[\exp\left\{\kappa 
\left|XY\right|\right\} \right]
&= \int_{-\infty}^{\infty}\dint x \frac{1}{\sqrt{2\pi}} e^{-x^2/2} \E\left[\exp\left\{\kappa |x| \cdot |Y| \right\} \right] \\
&= \int_{-\infty}^{\infty}\dint x \frac{1}{\sqrt{2\pi}} e^{-x^2/2} \cdot 2\exp\left\{\frac{1}{2}\kappa^2 x^2 \right\} \Phi(\kappa |x|)  \\
&= 2 \int_{-\infty}^{\infty}\dint x \frac{1}{\sqrt{2\pi}} \exp\left\{-\frac{1}{2}(1-\kappa^2) x^2 \right\} \Phi(\kappa |x|)  \\
&= 4 \int_{0}^{\infty}\dint x \frac{1}{\sqrt{2\pi}} \exp\left\{-\frac{1}{2}(1-\kappa^2) x^2 \right\} \Phi(\kappa x) \\
&\dfn F(\kappa),
\end{align}
where $\Phi(\cdot)$ is the normal cumulative distribution function.
Note that $F(\kappa) < \infty$ if and only if $\kappa \in (-1,1)$. To this end,
\begin{align}
\P\left[ \sum_{k=1}^{n} 
\left|X_{k} Y_{k} \right| \geq n\lambda \sigma \right] 
&\leq \inf_{\kappa \in (0,1)} e^{-n\kappa \lambda \sigma} \cdot \prod_{k=1}^{n} F(\kappa)\\
&= \inf_{\kappa \in (0,1)} e^{-n\kappa \lambda \sigma + n\log F(\kappa)}  \\
&= \exp\left\{-n \cdot \sup_{\kappa \in (0,1)} [\kappa \lambda \sigma - \log F(\kappa)]   \right\} \\
&\dfn \exp\left\{-n \cdot G(\lambda \sigma) \right\}.
\end{align}
For the right-hand probability in \eqref{ref91}, we utilize Chernoff's bound. Concretely, starting with 
\begin{align}
\P\left[ \sum_{i=1}^{n}X_{i}^2 \leq n\sigma \right]
&= \P\left[\exp\left\{-\tau \sum_{i=1}^{n}X_{i}^2 \right\} \geq e^{-n\tau\sigma} \right] \\
&\leq e^{n\tau\sigma} \cdot \E\left[\exp\left\{-\tau \sum_{i=1}^{n}X_{i}^2 \right\} \right] \\
&= e^{n\tau\sigma} \cdot (1+2\tau)^{-n/2},
\end{align}
we obtain the familiar bound
\begin{align}
\P\left[ \sum_{i=1}^{n}X_{i}^2 \leq n\sigma \right]
&\leq \inf_{\tau>0}
\exp\left\{n\tau\sigma -\frac{n}{2}\log(1+2\tau) \right\} \\
&= \exp\left\{-n \cdot \sup_{\tau>0} \left[\frac{1}{2}\log(1+2\tau) - \tau\sigma\right] \right\} \\
&= \exp\left\{-\frac{n}{2} \left[\sigma - \log(\sigma) - 1\right]\right\}.
\end{align}
Now,
\begin{align}
\P\left[ \calF_n^{\mbox{\scriptsize c}}(\sigma) \right] 
&= \P\left[ \calA_n^{\mbox{\scriptsize c}}(\sigma) \cup \calB_n^{\mbox{\scriptsize c}}(\sigma) \right] \\
&\leq \P\left[ \calA_n^{\mbox{\scriptsize c}}(\sigma) \right] + \P\left[  \calB_n^{\mbox{\scriptsize c}}(\sigma) \right] \\
&\leq 2\exp\left\{-\frac{n}{2} \left[\sigma - \log(\sigma) - 1\right]\right\} \\
&\dfn 2\exp\left\{-n \cdot J(\sigma)\right\}.
\end{align}
Let us continue from \eqref{ref91}. We optimize over $\sigma \in (0,1)$ and arrive at 
\begin{align}
\P\left[ \text{BC}(\bU, \bV) \geq \lambda \right]
&\leq \inf_{\sigma \in (0,1)} \left\{ \P\left[ \sum_{k=1}^{n} 
\left|X_{k} Y_{k} \right| \geq n\lambda \sigma \right] + \P\left[ \calF_n^{\mbox{\scriptsize c}}(\sigma) \right] \right\} \\
&\leq \inf_{\sigma \in (0,1)} 
\left\{\exp\left\{-n \cdot G(\lambda \sigma) \right\} + 2\exp\left\{-n \cdot J(\sigma)\right\} \right\} \\
&\leq \inf_{\sigma \in (0,1)} 
4 \exp\left\{-n \cdot \min\{G(\lambda \sigma), J(\sigma)\} \right\} \\
&= 4 \exp\left\{-n \cdot \sup_{\sigma \in (0,1)} \min\{G(\lambda \sigma), J(\sigma)\} \right\} \\
\label{ref92}
&\dfn 4\exp\left\{-n \cdot L(\lambda) \right\}.
\end{align}

Upper-bounding \eqref{ref90} with \eqref{ref92} and minimizing over $\lambda \in (0,1)$ provides
\begin{align}
\E\left[ 
\text{BC}(\bP_{m}, \bP_{m'})^{nr_n/\rho} \right] 
&\leq \inf_{\lambda \in (0,1)}
\left\{ \lambda^{nr_n/\rho} + 4\exp\left\{-n \cdot L(\lambda) \right\} \right\} \\
&\leq 8 \inf_{\lambda \in (0,1)} \exp\left\{-n \cdot \min\left[\frac{r_n}{\rho}\log\left(\frac{1}{\lambda}\right), L(\lambda) \right] \right\} \\
\label{ref94}
&\dfn 8 \exp\left\{-n \cdot S(r_n,\rho) \right\},
\end{align}
where
\begin{align}
S(r_n,\rho) := \sup_{\lambda \in (0,1)} \min\left[\frac{r_n}{\rho}\log\left(\frac{1}{\lambda}\right), L(\lambda)\right].
\end{align}
We now upper-bound \eqref{ref93} with \eqref{ref94} and find that for any $m \in [M]$
\begin{align}
\E\left[P_{e|m}^{1/\rho}\right] 
&\leq 8 (e^{nR}-1) \exp\left\{-n \cdot S(r_n,\rho) \right\} \\
&\leq 8 \exp\left\{-n \cdot (S(r_n,\rho)-R) \right\}.
\end{align}

Given this upper bound on the fractional moment of the error probability, we may now prove the existence of an expurgated code. According to Markov's inequality, it follows that
\begin{align}
\P\left[\frac{1}{M}\sum_{m=1}^{M} P_{e|m}^{1/\rho} > 16 \exp\left\{-n \cdot (S(r_n,\rho)-R) \right\} \right] \leq \frac{1}{2},
\end{align}
which means that there exists a code with
\begin{align}
\frac{1}{M}\sum_{m=1}^{M} P_{e|m}^{1/\rho} \leq 16 \exp\left\{-n \cdot (S(r_n,\rho)-R) \right\}.
\end{align}
We conclude that there exists a code $\calC_n$ with $M/2$ codewords for which 
\begin{align}
\max_{m} P_{e|m}^{1/\rho} \leq 16 \exp\left\{-n \cdot (S(r_n,\rho)-R) \right\},
\end{align}
and so,
\begin{align}
\max_{m} P_{e|m} \leq 16^{\rho} \exp\left\{-n \cdot \rho \cdot (S(r_n,\rho)-R) \right\},
\end{align}
thus,
\begin{align}
\liminf_{n \to \infty} -\frac{1}{n} \log \max_{m} P_{e|m} \geq \rho \cdot (S(r_n,\rho)-R).
\end{align}
Since it holds for any $\rho \geq 1$, the negative exponential rate of the error probability can be bounded as 
\begin{align}
\liminf_{n \to \infty} -\frac{1}{n} \log \max_{m} P_{e|m} \geq \sup_{\rho >1} \{\rho \cdot (S(r_n,\rho)-R)\},
\end{align}
and the proof of Theorem \ref{Theorem_Expurgated} is now complete.

\section{Conclusion} \label{Section6}
In this paper, we considered the error exponent and capacity of a channel whose input is a frequency vector of objects in a pool with unlimited resolution, and its output is samples of objects from that input frequency. We addressed the regime in which the number of samples scales with the input dimension (number of objects). As this channel is not memoryless, we derived a random coding bound from the first principles, and also used it to derive an achievable bound the capacity. We compared the unlimited input- resolution bound to capacity bounds derived previously for limited resolution. This comparison is currently partial, and an exact characterization of the capacity as a function of the input resolution, with exact limit at infinite resolution is an interesting open problem. From a technical perspective, our bound was derived for a restricted Dirichlet distribution, and relaxing this restriction might aid in achieving this goal. In addition, we derived an expurgated bound, which resulted improved bounds at low rates. It is of interest to further refine this bound, and similarly to above, this can be achieved by addressing the technical challenges involved in general input distributions. 
It is also of interest to derive upper bounds on the error exponent, which will hopefully show tightness of the exponents in some regimes. As the channel under consideration is non-standard, this might require novel ideas, beyond the techniques used to prove such bounds for classical channels.  Finally, we assumed that the  identification of the sampled object is noiseless, and it would be interesting to evaluate the effect of identification noise on the capacity and error exponent of the channel.
\appendices{\numberwithin{equation}{section}}

\section{Proof of Lemma \ref{LD_lemma} \label{appendix_lemma_proof}} 
Let $\{\rho_n\}_{n\geq 1}$ be any positive sequence. Consider the following bound
\begin{align}
\P\lb D\left(\frac{\bZ}{nr_n}\middle\|\bp\right) \geq \rho_n \rb
&= \P\lb \sum_{i=1}^{n} \frac{Z_i}{nr_n} \log \frac{Z_{i}}{nr_n p_i} \geq \rho_n \rb \\
&= \P\lb \sum_{i=1}^{n} Z_i \log \frac{Z_{i}}{p_i} - \sum_{i=1}^{n} Z_i \log (nr_n) \geq nr_n \rho_n \rb \\
&= \P\lb \sum_{i=1}^{n} Z_i \log \frac{Z_{i}}{p_i}  \geq nr_n [\rho_n + \log (nr_n)] \rb \\
&= \P\lb \prod_{i=1}^{n} \la\frac{Z_{i}}{p_i}\ra^{Z_i}  \geq \exp\lc nr_n [\rho_n + \log (nr_n)] \rc \rb \\
\label{ref2}
&\leq \frac{\E\lb \prod_{i=1}^{n} \la\frac{Z_{i}}{p_i}\ra^{Z_i}\rb}{\exp\lc nr_n [\rho_n + \log (nr_n)] \rc},
\end{align}
where \eqref{ref2} follows from Markov's inequality.

Recall the following Stirling bounds on $n!$:
\begin{equation}\label{Stirling}
\begin{aligned}
n! &\geq \sqrt{2\pi\max\left\{n,\tfrac{1}{2\pi}\right\}} \left(\frac{n}{e}\right)^{n} \geq \sqrt{2\pi n} \left(\frac{n}{e}\right)^{n},~~n \in \{0,1,2,\ldots\}, \\
n! &\leq \sqrt{2\pi en} \left(\frac{n}{e}\right)^{n},~~n \in \{1,2,\ldots\},
\end{aligned}
\end{equation}
where the maximization in the lower bound has been introduced to provide the exact factorial value for $n=0$.

We then evaluate the expectation in \eqref{ref2} as follows:
\begin{align}
\E\lb \prod_{i=1}^{n} \la\frac{Z_{i}}{p_i}\ra^{Z_i}\rb
&= \sum_{\bz \in \calS_n} P_{\bZ}(\bz) \prod_{i=1}^{n} \la\frac{z_{i}}{p_i}\ra^{z_i} \\
&= \sum_{\bz \in \calS_n} \frac{(nr_n)!}{z_1! \cdots z_n!} p_1^{z_1} \cdots p_n^{z_n} \prod_{i=1}^{n} \la\frac{z_{i}}{p_i}\ra^{z_i} \\
&= \sum_{\bz \in \calS_n} \frac{(nr_n)!}{z_1! \cdots z_n!} \prod_{i=1}^{n} z_{i}^{z_i} \\
\label{ref20}
&\leq \sum_{\bz \in \calS_n} \frac{\sqrt{2\pi e  nr_n}e^{-nr_n}(nr_n)^{nr_n}}{\prod_{i=1}^{n} \sqrt{2\pi \max\{z_i,\frac{1}{2\pi}\}} e^{-z_i} z_{i}^{z_i}} \prod_{i=1}^{n} z_{i}^{z_i} \\
&= \sum_{\bz \in \calS_n} \frac{\sqrt{2\pi e nr_n}(nr_n)^{nr_n}}{\prod_{i=1}^{n} \sqrt{2\pi \max\{z_i,\frac{1}{2\pi}\}} }  \\
\label{ref21}
&= \sqrt{2\pi e nr_n}(nr_n)^{nr_n} (2\pi)^{-n/2} \sum_{\bz \in \calS_n} \prod_{i=1}^{n}  \frac{1}{\sqrt{\max\{z_i,\frac{1}{2\pi}\}}},
\end{align}
where \eqref{ref20} follows from \eqref{Stirling}.
Let $U_1,\ldots,U_n$ be independent and identically distributed random variables, where $U_i$ is uniformly distributed over $\{0,1,\ldots,nr_n\}$.
The sum in \eqref{ref21} is upper-bounded in the following manner: \begin{align}
&\sum_{(u_1,\ldots,u_n) \in \calS_n} \prod_{i=1}^{n}  \frac{1}{\sqrt{\max\{u_i,\frac{1}{2\pi}\}}} \nn \\
&= (nr_n+1)^n \sum_{u_1=0}^{nr_n} \cdots \sum_{u_n=0}^{nr_n} \la \frac{1}{nr_n+1}\ra^n \I\{u_1+\ldots+u_n=nr_n\} \prod_{i=1}^{n}  \frac{1}{\sqrt{\max\{u_i,\frac{1}{2\pi}\}}} \\ 
&= (nr_n+1)^n 
\E \lb \I\{U_1+\ldots+U_n=nr_n\} \prod_{i=1}^{n}  \frac{1}{\sqrt{\max\{U_i,\frac{1}{2\pi}\}}} \rb \\
\label{ref22}
&\leq (nr_n+1)^n 
\left(\E \lb \left(\I\{U_1+\ldots+U_n=nr_n\}\right)^{p} \rb \right)^{1/p} \cdot
\left(\E \lb \left(\prod_{i=1}^{n}  \frac{1}{\sqrt{\max\{U_i,\frac{1}{2\pi}\}}} \right)^{q} \rb \right)^{1/q} \\
&= (nr_n+1)^n 
\left(\P \lb U_1+\ldots+U_n=nr_n   \rb \right)^{1/p} \cdot
\left(\E \lb \prod_{i=1}^{n}  \frac{1}{ \left( \max\{U_i,\frac{1}{2\pi}\} \right)^{q/2} }  \rb \right)^{1/q} \\
\label{ref23}
&= (nr_n+1)^n 
\left(\P \lb U_1+\ldots+U_n=nr_n   \rb \right)^{1/p} \cdot \prod_{i=1}^{n}
\left( \E \lb \frac{1}{ \left( \max\{U_i,\frac{1}{2\pi}\} \right)^{q/2} }  \rb \right)^{1/q},
\end{align}
where \eqref{ref22} is due to H\"older's inequality with $p,q >1$ and $\frac{1}{p}+\frac{1}{q}=1$. 
The probability in \eqref{ref23} is bounded as 
\begin{align}
\P[U_1+\ldots+U_n=nr_n]
&= \frac{|\calS_n|}{(1+nr_n)^n} \\
&= \frac{1}{(1+nr_n)^n} \binom{nr_n+n-1}{n-1} \\
\label{ref10}
&\leq \frac{1}{(1+nr_n)^n} \binom{nr_n+n}{n} \\
\label{ref14}
&\leq \frac{1}{(1+nr_n)^n} \sqrt{\frac{e(1+ r_n)}{2\pi nr_n}} \exp \lc n(1+r_n) H_{2}\la \frac{1}{1+r_n} \ra \rc \\
\label{ref24}
&\leq \frac{1}{(1+nr_n)^n} \exp \lc n \Psi(r_n) \rc, 
\end{align}
where \eqref{ref10} follows from the identity 
\begin{align}
\binom{n}{k} = \binom{n-1}{k-1} + \binom{n-1}{k},    
\end{align}
which holds for all integers $n,k$ such that $1 \leq k < n$, and \eqref{ref14} is due to \eqref{Stirling}. 
The expectation in \eqref{ref23} is bounded as
\begin{align}
\E \lb \frac{1}{\left(\max\{U_i,\frac{1}{2\pi}\}\right)^{q/2} } \rb
&= \frac{1}{nr_n+1} \sum_{m=0}^{nr_n} \frac{1}{\left(\max\{m,\frac{1}{2\pi}\}\right)^{q/2} } \\
&= \frac{1}{nr_n+1} \left[(2\pi)^{q/2} + \sum_{m=1}^{nr_n} \frac{1}{m^{q/2}} \right] \\
&\leq \frac{1}{nr_n+1} \left[(2\pi)^{q/2} + \sum_{m=1}^{\infty} \frac{1}{m^{q/2}} \right] \\
\label{ref25}
&= \frac{1}{nr_n+1} \left[(2\pi)^{q/2} + \zeta\left(\frac{q}{2}\right) \right],
\end{align}
where $\zeta(s)$ is Riemann's zeta function and we assume that $q>2$. 
Substituting \eqref{ref24} and \eqref{ref25} back into \eqref{ref23} yields 
\begin{align}
&\sum_{(u_1,\ldots,u_n) \in \calS_n} \prod_{i=1}^{n}  \frac{1}{\sqrt{\max\{u_i,\frac{1}{2\pi}\}}} \nn \\
&~~~\leq (nr_n+1)^n \frac{1}{(nr_n+1)^{n/p}} \exp \lc \frac{n}{p} \Psi(r_n) \rc 
\cdot \frac{1}{(nr_n+1)^{n/q}} \left[(2\pi)^{q/2} + \zeta\left(\frac{q}{2}\right) \right]^{n/q} \\
\label{ref26}
&~~~= \exp \lc \frac{n}{p} \Psi(r_n) \rc \cdot \left[(2\pi)^{q/2} + \zeta\left(\frac{q}{2}\right) \right]^{n/q}.
\end{align}
Upper-bounding \eqref{ref21} using \eqref{ref26} yield
\begin{align}
\E\lb \prod_{i=1}^{n} \la\frac{Z_{i}}{p_i}\ra^{Z_i}\rb
&\leq \sqrt{2\pi e nr_n}(nr_n)^{nr_n} (2\pi)^{-n/2} \cdot \exp \lc \frac{n}{p} \Psi(r_n) \rc \cdot \left[(2\pi)^{q/2} + \zeta\left(\frac{q}{2}\right) \right]^{n/q} \\
&= \sqrt{2\pi e nr_n}(nr_n)^{nr_n} \cdot \exp \lc \frac{n}{p} \Psi(r_n) \rc \cdot \left[(2\pi)^{-q/2}(2\pi)^{q/2} + (2\pi)^{-q/2} \zeta\left(\frac{q}{2}\right) \right]^{n/q} \\
&= \sqrt{2\pi e nr_n}(nr_n)^{nr_n} \cdot \exp \lc \frac{n}{p} \Psi(r_n) \rc \cdot \left[1 + (2\pi)^{-q/2} \zeta\left(\frac{q}{2}\right) \right]^{n/q} \\
&= \sqrt{2\pi e nr_n}(nr_n)^{nr_n} \cdot \exp \lc n\left(1-\frac{1}{q}\right) \Psi(r_n) + \frac{n}{q}\log\left[1 + (2\pi)^{-q/2} \zeta\left(\frac{q}{2}\right) \right] \rc, 
\end{align}
which is finite for any $q>2$, and thus
\begin{align}
&\E\lb \prod_{i=1}^{n} \la\frac{Z_{i}}{p_i}\ra^{Z_i}\rb \nn \\
&\leq \sqrt{2\pi e nr_n}(nr_n)^{nr_n} \cdot \exp \left\{ n \cdot \inf_{q>2} \left\{ \left(1-\frac{1}{q}\right) \Psi(r_n) + \frac{1}{q}\log\left[1 + (2\pi)^{-q/2} \zeta\left(\frac{q}{2}\right) \right] \right\} \right\} \\
\label{ref27}
&= \sqrt{2\pi e nr_n}(nr_n)^{nr_n} \cdot \exp \left\{ n \cdot \Delta(r_n) \right\}.
\end{align}
Upper-bounding \eqref{ref2} with \eqref{ref27} finally provides
\begin{align}
\P\lb D\left(\frac{\bZ}{nr_n}\middle\|\bp\right) \geq \rho_n \rb
&\leq \frac{\E\lb \prod_{i=1}^{n} \la\frac{Z_{i}}{p_i}\ra^{Z_i}\rb}{\exp\lc nr_n [\rho_n + \log (nr_n)] \rc} \\
&\leq \sqrt{2\pi e nr_n} \cdot \frac{(nr_n)^{nr_n} \exp \lc n \cdot \Delta(r_n) \rc }{\exp\lc nr_n [\rho_n + \log (nr_n)] \rc } \\
&= \sqrt{2\pi e nr_n} \cdot \exp \lc -n (r_n \rho_n-\Delta(r_n)) \rc.
\end{align}

For $\mu>0$, let us now choose 
\begin{align}
\rho_n = \frac{\Delta(r_n) + \mu}{r_n}
\end{align}
which yields
\begin{align}
\P\lb D\left(\frac{\bZ}{nr_n}\middle\|\bp\right) \geq \frac{\Delta(r_n) + \mu}{r_n} \rb
&\leq \sqrt{2\pi e nr_n} e^{-n\mu} ,
\end{align}
and this completes the proof of Lemma \ref{LD_lemma}.

\bibliographystyle{plain}
\bibliography{DNA_concentration.bib}

\end{document}